\documentclass[aps,epsfig]{revtex4}

\usepackage{amssymb}
\usepackage[dvips]{graphicx} 

\newcommand{\fix}{\Phi(\mathbf{x})}
\newcommand{\fiLx}{\Phi_{\rm L}(\mathbf{x})}
\newcommand{\fiLxsquare}{\Phi^2_{\rm L}(\mathbf{x})}

\newcommand{\beq}{\begin{equation}}
\newcommand{\eeq}{\end{equation}}
\newcommand{\beqarr}{\begin{eqnarray}}
\newcommand{\eeqarr}{\end{eqnarray}}

\newcommand{\almone}{a_{\ell_1 m_1}}

\newcommand{\almtwo}{a_{\ell_2 m_2}}

\newcommand{\almthree}{a_{\ell_3 m_3}}

\newcommand{\Gaunt}{\mathcal{G}_{\ell_1 \ell_2 \, \ell_3}^{m_1 m_2 m_3}}

\newcommand{\Bis}{B_{\ell_1 \ell_2 \ell_3}^{m_1 m_2 m_3}}
\newcommand{\Avbis}{B_{\ell_1 \ell_2 \ell_3}}
\newcommand{\Redbis}{b_{\ell_1 \ell_2 \ell_3}}
\newcommand{\Redbisloc}{b^{\rm local}_{\ell_1 \ell_2 \ell_3}}
\newcommand{\Redbiseq}{b^{\rm equil}_{\ell_1 \ell_2 \ell_3}}
\newcommand{\fnlloc}{f_{\rm NL}^{\rm local}}
\newcommand{\fnleq}{f_{\rm NL}^{\rm equil}}

\newcommand{\fnl}{f_{\rm NL}}
\newcommand{\fnlest}{\hat{f}_{\rm NL}}
\newcommand{\A}{A_S}

\begin{document}

\author{Michele Liguori$^1$ and Antonio Riotto$^2,^3$}

\affiliation{$^1$Department of Applied Mathematics and Theoretical Physics, Centre for
Mathematical Sciences, University of Cambridge, Wilberfoce Road, Cambridge, CB3 0WA, United
Kingdom}

\affiliation{$^2$INFN, sezione di Padova, via Marzolo 8, I-35131, Padova,
  Italy} 
\affiliation{$^3$CERN, Theory Division, CH-1211 Geneva 23, Switzerland}

\title{Impact of Uncertainties in the Cosmological Parameters on\\
  the Measurement of Primordial non-Gaussianity}

\begin{abstract}
\noindent We study the impact of cosmological parameters' uncertainties on
estimates of the primordial NG parameter $\fnl$ in local and
equilateral models of non-Gaussianity. We show
that propagating these errors increases the $\fnl$ $1\sigma$ uncertainty
by a term ${\delta \fnlloc / \fnlloc} \simeq 16 \%$ for WMAP and  
${\delta \fnlloc / \fnlloc} \simeq 5 \%$ for Planck in the local case, whereas for
equilateral configurations the correction term are  ${\delta \fnleq / \fnleq}
\simeq 14 \%$ and ${\delta \fnleq / \fnleq } \simeq 4\%$, respectively. If we 
assume for $\fnlloc$ a central value $\langle \fnlloc \rangle \simeq
60$, according to recent WMAP 5-years estimates, we obtain for Planck a final
correction $\Delta \fnlloc \simeq 3$. Although not dramatic, this
correction is at the level of the expected estimator uncertainty for Planck, and
should then be taken into account when quoting the significance of an
eventual future detection.
In current estimates of $\fnl$ the cosmological parameters
are held fixed at their best-fit values. We finally note that the
impact of uncertainties in the cosmological
parameters on the final $\fnl$ error bar would become totally negligible if
the parameters were allowed to vary in the analysis and then
marginalized over.
\end{abstract}

\maketitle

\section{Introduction}

\noindent Research of primordial non-Gaussianity in the Cosmic Microwave
Background (CMB) is an important field of cosmology today. 
In a recent work, Yadav and Wandelt \cite{YadavWandelt} claim 
a $3 \sigma$ detection of a large NG signal in the 
WMAP 3-years data.
The following WMAP 5-years analysis produced similar results but
showed a reduction of statistical significance
from about $3$ to $2 \sigma$ \cite{WMAP5}. 
Yadav and Wandelt estimate, using WMAP 3-years data, is
\beq
27 \leq \fnlloc \leq 147  \;\;  (95\% \, {\rm c.l.}) \; .
\eeq
The WMAP 5-years estimate is:
\beq
-9 \leq \fnlloc \leq 111  \;\;  (95\% \, {\rm c.l.}) \; ,
\eeq
where $\fnl$ is a dimensionless parameter defining the strength of
primordial non-Gaussianity (NG) \cite{NGreview}, and the superscript 
''local'' indicates that
we are considering a primordial NG curvature perturbation of
the form 
\beq
\fix = \fiLx + \fnlloc \left(\fiLxsquare - \langle \fiLxsquare \rangle
\right) \; ,
\eeq
where $\fix$ is the total gravitational potential, $\fiLx$ is the gravitational potential
computed at the linear level and $\langle\cdots \rangle$ stands for the 
ensemble average.  
The reason for calling this a local NG model
relies in the fact that the NG part of the primordial
curvature perturbation is a local functional of the Gaussian part.
The local shape of NG arises from standard single-field
slow-roll inflation \cite{standard} as well as from alternative inflationary 
scenarios for the generation of primordial perturbations, like 
the curvaton \cite{curvaton} or inhomogenous (pre)reheating models \cite{gamma1}, or even from
alternatives to inflation, such as ekpyrotic and cyclic models 
\cite{ek}. Other models, such as DBI inflation \cite{dbi} and ghost
inflation \cite{ghost},
predict a different kind of primordial NG, called 
  "equilateral", because the three point function for this kind of
NG is peaked on equilateral configurations, in which the
lengths of the three wavevectors forming a triangle in Fourier space are
equal \cite{shape}. This second form of NG is characterised by the
parameter $\fnleq$, which defines the amplitude of the equilateral
triangles. In the following we will sometimes write the amplitude
of NG simply as $\fnl$, without any superscript. When we do so,
we mean that our conclusions apply to both the local and the equilateral
case, with no need for distinction.

Standard single field inflation predicts
$\fnlloc \sim 10^{-2}$ at the end of inflation \cite{standard}  (and therefore a final value 
$\fnlloc \sim $ unity  after general relativistic second-order perturbation 
effetcs are taken into account \cite{second}). It is thus clear
that large central values of $\fnlloc$, like those obtained in the
above mentioned analyses, are going to rule out the simplest scenarios 
of inflation as viable models of the Early Universe. On the other
hand, the low statistical significance of the final WMAP 5-years
result makes any conclusion premature at this stage. With its high
angular resolution and sensitivity Planck will allow to significantly
improve the statistical estimate of $\fnlloc$, reducing the final error
bars from the present $\Delta \fnlloc \simeq 30$ to a final value of
$\Delta \fnlloc \simeq 5$ \cite{KomatsuSpergel} 
thus allowing a many $\sigma$ detection of non-Gaussianity if the
present large central values of $\fnlloc$ were to be confirmed. An
eventual detection of a large $\fnl$ by Planck would not  however
automatically imply that the observed non-Gaussianity is primordial in
origin. A number of effects can produce a {\em spurious} NG
signal that can bias the final estimate. The most relevant examples of
NG contaminants are probably given by diffuse foregrounds emission,
unresolved point sources contamination, NG noise. Both the
analyses by Yadav and Wandelt and by the WMAP team consider all these
effects and conclude that they do not significantly affect the $\fnlloc$
measurement from WMAP data. The picture gets however more complicated
when we consider future Planck data.  It has been recently shown 
that several effects that
are not important in the analysis of WMAP data become no longer negligible
with Planck. For example, Serra and Cooray \cite{SerraCooray} studied
NG contributions arising 
from several secondary sources, and concluded that effects such as the
cross-correlations SZ-lensing and ISW-lensing produce a bias in the
estimate of $\fnlloc$ which is at the level of the expected estimator
variance at Planck angular resolution. Analogous conclusions have been reached
by Babich and Pierpaoli \cite{BabichPierpaoli} for the
cross correlations of density and lensing magnification of radio and SZ
point sources with the ISW effect. Note that all these effects are
unimportant for present analyses both because they involve higher
multipoles than those reached at the WMAP angular resolution, and
because they produce a bias $\Delta \fnlloc \sim 1$, thus much smaller than
the WMAP sensitivity to $\fnlloc$, which is $\Delta \fnlloc \sim 30$. 
However, the much higher angular resolution
achieved by Planck and an expected predicted sensitivity on $\fnlloc$
given by $\Delta \fnlloc \sim 5$ (and $\Delta \fnlloc \sim 3$ if
polarisation data are included in the analysis) 
will make the above mentioned sources of NG  
contamination no longer negligible in the future.
In other words, the same 
nice properties that make Planck more sensitive to the detection of a primordial
NG signal (i.e. high angular resolution and sensitivity)
make it in fact also much more sensitive to the observation of
NG contaminants and require a very careful investigation of
all the potential sources of bias in the estimate of $\fnl$.

In this paper we will consider another  potential source of uncertainty in the
detection of NG, namely
the propagation of the uncertainties 
in the cosmological parameters on the measured value of $\fnl$. This
effect can be summarised as follows: the
estimator usually employed to measure $\fnl$ assumes a given
underlying cosmological model obtained by fixing the cosmological
parameters at their best-fit values (obtained from the two-point CMB likelihood
analysis of the experiment under exam). However the cosmological
parameter estimates are characterised by uncertainties that should be
propagated into the $\fnl$ estimate in order to accurately quote the
final error bars. The uncertainties in the
cosmological parameters can be safely neglected as long as they are
much smaller than the variance of the NG estimator. While this works well 
for WMAP, it is a priori unclear whether it is still a good approximation 
for Planck. 
 
The paper is structured as follow: in
section \ref{sec:bias} we will describe the $\fnl$ estimator
commonly employed in the analysis and study in detail the effect of 
cosmological parameters uncertainties on this estimator. After
deriving the error propagation formulae (\ref{eqn:bias}) and
(\ref{eqn:propagation}) we will apply them to obtain analytical and
numerical estimates of the expected $\fnl$ uncertainty for WMAP and
Planck, both in the local and equilateral case.
In section \ref{sec:Fisher} we will then consider the possibility of 
applying a more complex analysis in which $\fnl$ is estimated by
firstly allowing the cosmological parameters to vary and then by marginalising over
them, rather than by fixing the cosmological model. In
this case we 
adopt a Fisher matrix approach to propagate the parameter uncertainties
on the final predicted $\Delta \fnl$. We will finally discuss
our results and draw our conclusion in section \ref{sec:Conclusions}.

\section{Errors propagation}\label{sec:bias}
\noindent
The estimate of $\fnl$ from CMB data are usually obtained from
measurements of the three point function in harmonic space,
called the angular bispectrum and defined as:
\beq
\Bis \equiv \left\langle a_{\ell_1}^{m_1} a_{\ell_2}^{m_2}
a_{\ell_3}^{m_3} \right\rangle \; .
\eeq
Due to rotational invariance of the CMB sky the angular bispectrum can
be written as:
\beq
\Bis = \Gaunt \Redbis \; ,
\eeq
where $\Gaunt$ is the Gaunt integral:
\beq
\Gaunt = 
 \sqrt{\frac{(2 \ell_1 +1)(2 \ell_2 + 1)(2 \ell_3 + 1)}{4 \pi}}
             \left( \begin{array}{ccc} 
	     \ell_1 & \ell_2 & \ell_3 \\
	       0      &  0     &  0    
	           \end{array} \right)
             \left( \begin{array}{ccc} 
             \ell_1 & \ell_2 & \ell_3 \\
               m_1      &  m_2     &  m_3    
             \end{array} \right)  \, ,
\eeq
while $\Redbis$ is called the reduced bispectrum and contains all the
relevant physical information. Analytic formulae for both the local
and equilateral cases have been computed
\cite{KomatsuSpergel,Creminellietal}. The local reduced bispectrum can
be written as:
\beq\label{eqn:locbis}
\Redbisloc = 2 \fnlloc \int dr r^2 \alpha_{\ell_1}(r) \beta_{\ell_2}(r)
  \beta_{\ell_3}(r) + (2 \, {\rm perm}.) \; ,
\eeq
whereas the equilateral bispectrum is
\beq\label{eqn:eqbis}
\Redbiseq = 6 \fnleq \int dr r^2 \alpha_{\ell_1}(r) \beta_{\ell_2}(r)
\beta_{\ell_3}(r) + (2 \, {\rm perm}.) + \delta_{\ell_1}(r) \delta_{\ell_2}(r)
\delta_{\ell_3}(r) + \beta_{\ell_1}(r) \gamma_{\ell_2}(r)
\delta_{\ell_3}(r) + (5 \,
{\rm perm}.) \; .
\eeq
The functions
$\alpha_\ell(r)$,$\beta_\ell(r)$,$\gamma_\ell(r)$,$\delta_\ell(r)$
appearing in the previous formulae are defined as:
\beqarr\label{eqn:radialcoeff}
\alpha_\ell(r) & = & \frac{2}{\pi} \int dk k^2 \Delta_\ell(k)
j_\ell(kr)\;,  \nonumber \\
\beta_\ell(r) & = & \frac{2}{\pi} \int dk k^2 P_\Phi(k) \Delta_\ell(k)
j_\ell(kr)\; , \nonumber \\
\gamma_\ell(r) & = & \frac{2}{\pi} \int dk k^2 P_\Phi^{1/3}(k) \Delta_\ell(k)
j_\ell(kr)\; ,\nonumber \\ 
\delta_\ell(r) & = & \frac{2}{\pi} \int dk k^2 P_\Phi^{2/3}(k) \Delta_\ell(k)
j_\ell(kr) \; .
\eeqarr
In the previous set of formulae $\Delta_\ell(k)$ indicates the CMB
radiation transfer function and $P_\Phi(k)$ is the power spectrum of
primordial curvature perturbation. It is thus clear that the reduced
bispectrum will be dependent on the cosmological parameters.

The NG estimator
which is generally employed to analyse CMB data can be written as \cite{KSW1,KSW2}:
\beq
\hat{f}_{\rm NL} = \frac{1}{N} \sum_{\ell_1 \ell_2 \ell_3} 
\sum_{m_1  m_2 m_3} \Gaunt \frac{b_{\ell_1 \ell_2 \ell_3}}{C_{\ell_1} C_{\ell_2} C_{\ell_3}}
\almone \almtwo \almthree \; ,
\eeq
where $b_{\ell_1 \ell_2 \ell_3}$ is the analytical form of the
primordial reduced bispectrum for the model we are considering (i.e. either
local or equilateral), whereas $N$ is a normalisation factor designed
to produce unitary response when $\fnl = 1$:
\beq
N = \sum_{\ell_1 < \ell_2 < \ell_3} \frac{(2
  \ell_1+1)(2\ell_2+1)(2\ell_3+1)}{4 \pi} 
   \left( \begin{array}{ccc} 
	     \ell_1 & \ell_2 & \ell_3 \\
	       0      &  0     &  0    
	           \end{array} \right)^2
   \frac{(b^{\fnl = 1}_{\ell_1 \ell_2 \ell_3})^2}{C_{\ell_1} C_{\ell_2} C_{\ell_3}}
   \; .
\eeq
An additional linear term is added to the estimator when dealing with
anisotropic noise in the data. We will not include such term here as
it is not dependent on the cosmological parameters and thus does not affect
our estimates.
The estimated value of $\fnl$ is then obtained by correlating the
observed bispectrum with the theoretically expected one for a given
primordial shape (local or equilateral) and {\em given cosmological model}, 
and dividing by a suitable normalisation factor. Also the
normalisation will be dependent on the bispectrum shape and on the 
cosmological parameters, as both $b_{\ell_1 \ell_2 \ell_3}$ and
$C_\ell$ are. The $C_\ell$ and $b_{\ell_1 \ell_2 \ell_3}$ that enter
the estimator are calculated by assuming the best-fit cosmological
model for the experiment under consideration. However, the best-fit
cosmological parameters are characterised by uncertainties that
propagate to $\hat{f}_{\rm NL}$. The aim of this work is to analyse
this effect in detail and assess its significance for WMAP and Planck.
As a start, following \cite{Creminellietal} let us assume that the
cosmological model assumed in the calculation of $\Redbis$ and
$C_\ell$ is not the ``real'' one. Let us call the reduced bispectrum obtained from
the real cosmological parameters $\tilde{b}_{\ell_1 \ell_2
\ell_3}$. It is then easy to see that the average value of the
estimator will be \cite{Creminellietal}:
\beqarr
\langle \hat{f}_{\rm NL} \rangle & = & \frac{1}{N} \sum_{\ell_1 <
   \ell_2 <
   \ell_3}  \frac{(2
  \ell_1+1)(2\ell_2+1)(2\ell_3+1)}{4 \pi} 
   \left( \begin{array}{ccc} 
	     \ell_1 & \ell_2 & \ell_3 \\
	       0      &  0     &  0    
	           \end{array} \right)^2
   \frac{b^{\fnl = 1}_{\ell_1 \ell_2 \ell_3} \langle \almone \almtwo \almthree \rangle
       }{C_{\ell_1}C_{\ell_2} C_{\ell_3}} \nonumber \\
     & = &             \frac{1}{N} \sum_{\ell_1 < \ell_2 <
   \ell_3}  \frac{(2
  \ell_1+1)(2\ell_2+1)(2\ell_3+1)}{4 \pi} 
   \left( \begin{array}{ccc} 
	     \ell_1 & \ell_2 & \ell_3 \\
	       0      &  0     &  0    
	           \end{array} \right)^2
   \frac{b^{\fnl = 1}_{\ell_1 \ell_2 \ell_3} \tilde{b}_{\ell_1 \ell_2
       \ell_3}}{C_{\ell_1}C_{\ell_2} C_{\ell_3}} \; ,
\eeqarr 
throughout the paper we will use a superscript $\hat{}$ on $\fnl$
whenever we want to indicate a statistical estimate  
If we now want to estimate the bias $\delta \fnl$ due to the mismatch
between the assumed cosmological model and the real one, we can write:
\beqarr
\delta \fnlest \equiv \langle \hat{f}_{\rm NL} \rangle - \fnl & = & 
   \frac{1}{N} \sum_{\ell_1 < \ell_2 <\ell_3}  \frac{(2\ell_1+1)(2\ell_2+1)(2\ell_3+1)}{4 \pi} 
   \left( \begin{array}{ccc} 
	     \ell_1 & \ell_2 & \ell_3 \\
	       0      &  0     &  0    
	           \end{array} \right)^2
   \frac{b^{\fnl = 1}_{\ell_1 \ell_2 \ell_3} \tilde{b}_{\ell_1 \ell_2
   \ell_3}}{C_{\ell_1}C_{\ell_2} C_{\ell_3}} \nonumber \; ,\\
   & - & 
   \frac{\fnl}{N} \sum_{\ell_1 < \ell_2 <\ell_3}  \frac{(2\ell_1+1)(2\ell_2+1)(2\ell_3+1)}{4 \pi} 
   \left( \begin{array}{ccc} 
	     \ell_1 & \ell_2 & \ell_3 \\
             0      &  0     &  0    
	           \end{array} \right)^2
   \frac{b^{\fnl = 1}_{\ell_1 \ell_2 \ell_3} b^{\fnl =1}_{\ell_1 \ell_2
   \ell_3}}{C_{\ell_1} C_{\ell_2} C_{\ell_3}}
   \nonumber  \; ,\\
   & \simeq & 
      \frac{\fnl}{N} \sum_{\ell_1 < \ell_2 <\ell_3}  \frac{(2\ell_1+1)(2\ell_2+1)(2\ell_3+1)}{4 \pi} 
   \left( \begin{array}{ccc} 
	     \ell_1 & \ell_2 & \ell_3 \\
             0      &  0     &  0    
	           \end{array} \right)^2
   \frac{b^{\fnl = 1}_{\ell_1 \ell_2 \ell_3} \delta b_{\ell_1 \ell_2
   \ell_3}}{C_{\ell_1}C_{\ell_2} C_{\ell_3}} \; .
\eeqarr
In this last formula, $\delta b_{\ell_1 \ell_2
\ell_3}$ is the difference between the theoretical bispectra
computed for the ``real'' and ``assumed'' cosmological model; all the
quantities with a superscript $\sim$ are computed in the ``real''
cosmological model. If we
then have a cosmological parameter characterised by an uncertainty 
$\delta p$ we simply propagate this uncertainty on the $\fnl$
estimate as:
\beq\label{eqn:bias}
\delta  \fnlest  = \frac{\delta \fnlest}{\delta p} {\delta p} \simeq
      \left[ \frac{\fnl}{N} \sum_{\ell_1 < \ell_2 <\ell_3}  \frac{(2\ell_1+1)(2\ell_2+1)(2\ell_3+1)}{4 \pi} 
   \left( \begin{array}{ccc} 
	     \ell_1 & \ell_2 & \ell_3 \\
             0      &  0     &  0    
	           \end{array} \right)^2
   \frac{b^{\fnl = 1}_{\ell_1 \ell_2 \ell_3} \frac{\delta b_{\ell_1 \ell_2
   \ell_3}}{\delta p}}{C_{\ell_1}C_{\ell_2} C_{\ell_3}} \right] \delta
   p \; .
\eeq
The term in square brackets expresses
the derivative of $\fnlest$ with respect to the parameter $p$ as a
function of the derivative of the bispectrum with respect to  $p$. 
As an estimate of the uncertainty on the parameter $p$ we can use the
standard deviation and thus substitute $\delta p$, indicating a
general small variation of the parameter in the previous formula,
with $\sigma_p$, expressing its $1\sigma$ uncertainty. The formula
above then simply reads:
\beq
\sigma_{f_{\rm NL}} = \frac{\partial \fnlest}{\partial p} \sigma_p \; ;
\eeq        
this is the standard formula of error propagation. Note that here and in
the following equations,
we sometimes indicate uncertainties in a parameter
using the letter $\delta$ (e.g. $\delta p$), sometimes we use
$\sigma$, e.g. $\sigma_{\fnl}$. With $\delta$ we just generically
indicate a small variation in the parameter (possibly dependent on the
variation in one or more other parameters)
, whereas with $\sigma$ we
specifically refer to the standard deviation.
 
\begin{figure}[h]
\begin{center}
\includegraphics[height=0.6\textheight,width = 0.9\textwidth]{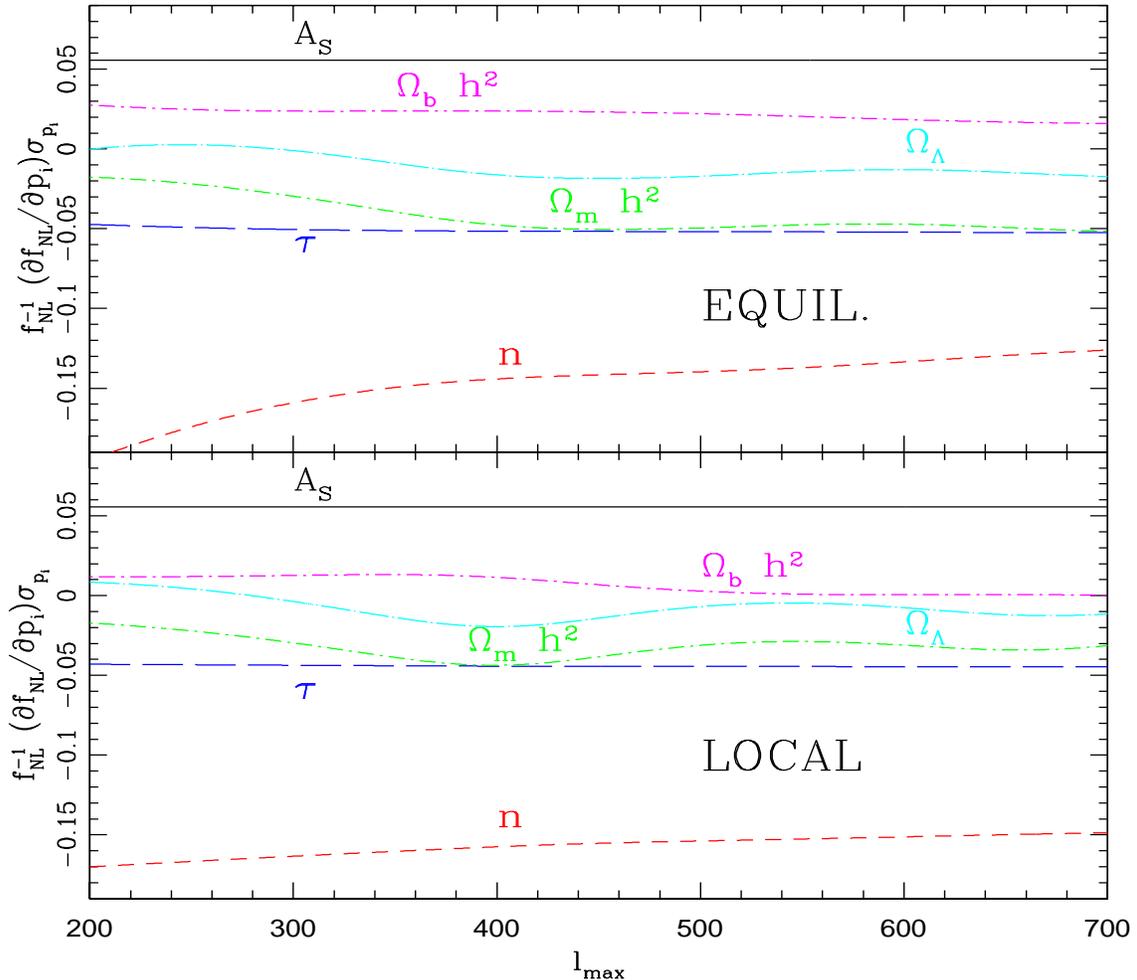}
\caption{Contribution of the different cosmological parameter
uncertainties to the final error in the estimate of $\fnlloc$ (lower
panel) and $\fnleq$  (upper panel) as a function of $\ell_{\rm max}$. The
quantity $({\partial \fnl / \partial p_i}) \sigma_{p_i}$ is plotted for each of the
six parameter in the model. In this figure we considered an experiment 
with the characteristics of WMAP.
}\label{fig:derivativeswmap}
\end{center}
\end{figure}

\begin{figure}[h]
\begin{center}
\includegraphics[height=0.6\textheight,width = 0.9\textwidth]{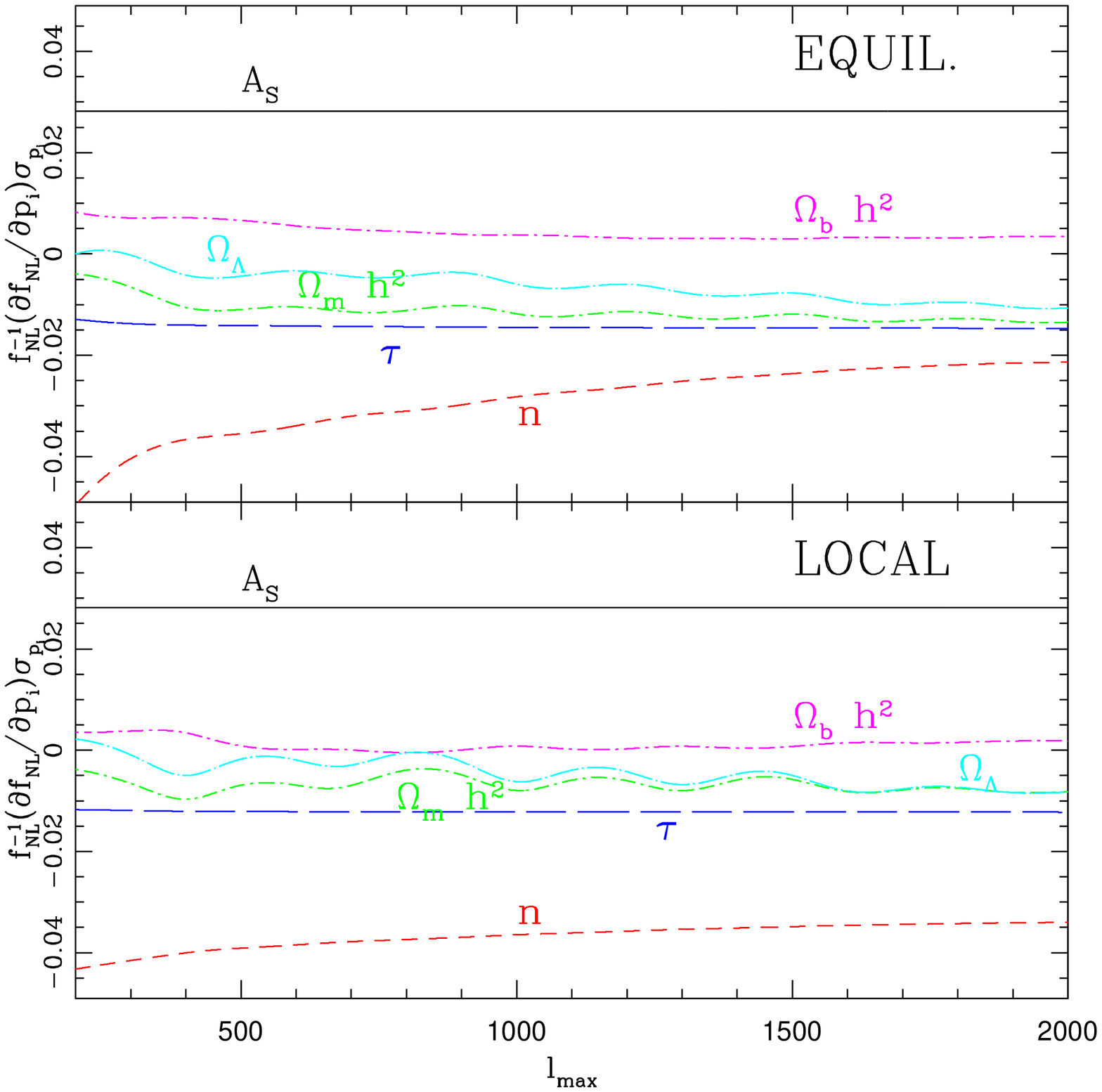}
\caption{Contribution of the different cosmological parameter
uncertainties to the final error in the estimate of $\fnlloc$ (lower
panel) and $\fnleq$ for an experiment 
with the characteristics of Planck.}\label{fig:derivativesplanck}
\end{center}
\end{figure}

In general we have a cosmological model defined by a set of parameters 
$\{p_i\}$. All these parameters are allowed to vary and can be correlated. 
The standard error propagation formula in this case becomes:
\beq\label{eqn:propagation}
\delta  \fnlest  = \sqrt{\sum_{ij}\left. \frac{\partial \fnl}{\partial
   p_i}\right|_{p_i = \bar{p}_i}
                    \left.\frac{\partial \fnl}{\partial
   p_j}\right|_{p_j=\bar{p}_j} {\rm Cov}(p_i,p_j)} 
\eeq  
where the average values $\bar{p}_i$ of the parameters and their  
covariance matrix ${\rm Cov}(p_i,p_j)$ are determined from a standard
CMB-likelihood
analysis, in which the anisotropies are assumed
Gaussian. 
In our analysis we considered a model characterised by the six
parameters $A_S,n_s,\tau,\omega_b = \Omega_b h^2,\omega_m
= \Omega_m h^2,\Omega_\Lambda\}$, respectively defining the amplitude of
curvature perturbation at $k_0 = $0.002/Mpc, the scalar spectral
index, the optical depth to reionization, the physical density of
baryons, the physical density of matter and the dark energy density.
We fixed these parameters at their maximum likelihood values from the
WMAP 5-year analysis and as an estimate of the covariance matrix we 
computed the Fisher matrix of the parameters. In the computation 
of the Fisher matrix we consider two cases: in the first case we use
the $41$, $61$ and $94$ Ghz frequency channels of WMAP (Q+V+W bands)
\cite{lambda} and
in the second we consider the combination of the $143$ and $217$ Ghz
frequency channels of Planck \cite{Bluebook}. For the numerical
details of the computation (e.g. choice of the step size for the
derivatives w.r.t. cosmological parameters) we closely followed the
methodology presented in \cite{HuTegmark}.
We then computed the CMB local and
equilateral bispectra numerically using formulae
(\ref{eqn:locbis}),(\ref{eqn:eqbis}),(\ref{eqn:radialcoeff}) and we
took two sided numerical derivatives to evaluate ${\partial B/ \partial
 p_i}$. For the steps of the derivatives we again followed the 
prescriptions of 
\cite{HuTegmark}. Our calculation shows that the $\fnl$ error bars get
relative corrections  ${\delta \fnlloc/ \fnlloc} \simeq 16.5 \%$ and 
${\delta \fnleq/ \fnleq} \simeq 14.5 \%$ for WMAP
in the local and equilateral case, while for Planck 
we have ${\delta \fnlloc / \fnlloc}
\simeq 5 \%$ and ${\delta \fnleq / \fnleq} \simeq 4.5 \%$. Before discussing the
significance of this correction let us try to understand this result
in a more intuitive way by employing some analytical approximations. 
From figures \ref{fig:derivativeswmap} and \ref{fig:derivativesplanck}
   we see that most
of the contribution to $\delta \fnl$ from error propagation comes 
from only 3 of the 6 considered parameters: $A_S$,$n_s$
and $\tau$ (this is in agreement with the results of
\cite{YadavWandelt}). We will then restrict the following 
simplified analysis to these three parameters.

The first step of this analysis is to get analytical
expressions for the derivatives of the bispectrum with respect to $A_S,n_S,\tau$.
From formulae (\ref{eqn:radialcoeff}) it can be
easily seen that ${\partial B / \partial A_S} = {2B / A_S}$. The
fractional variation in $\fnlest$ corresponding to a variation $\delta
A_S$ is then:
\beq
\frac{\delta \fnlest}{\delta A_S} \delta A_S = 2 \fnlest \frac{\delta A_S}{A_S} 
\eeq
The parameter $\tau$ defines the optical depth to
reionization and the effect of changing it can be described by
introducing a multiplicative factor $e^{-\tau}$ in front of the
radiation transfer function at high $\ell's$. Note that the radiation 
transfer functions appear in the definition of the bispectrum through
the functions $\alpha$, $\beta$, $\gamma$, $\delta$ defined in formula
(\ref{eqn:radialcoeff}). The bispectrum modes that give the largest 
contributions in the local case to the final signal-to-noise ratio are the so called
squeezed configuration, i.e. configuration where one
of the three $\ell's$ is much smaller than the other two, and
equilateral triangles in the other case (see
e.g. \cite{Creminellietal}). 
For this reason, in the local
case one of the three $\ell$'s will be super-horizon and the
corresponding transfer function will not show a multiplicative
factor $e^{-\tau}$ in front. In the equilateral case all modes are
sub-horizon in the important configurations. We can then write
$\tilde{b}_{\ell_1 \ell_2 \ell_3}^{\rm local} = \exp(-2\tau) \Redbisloc$
and $\tilde{b}_{\ell_1 \ell_2 \ell_3}^{\rm equil} = \exp(-3\tau)
\Redbiseq$. Substituting into equation (\ref{eqn:bias}) we find
$\delta \fnlloc \simeq -2 \fnlloc \delta \tau$ and 
$\delta \fnleq \simeq  -3 \fnleq \delta \tau$. Note
though that the parameter $\tau$ obtained from the $C_\ell$ likelihood analysis 
is degenerate with the amplitude of the spectrum of primordial curvature 
perturbations. In order to include this degeneration in our simplified
description, for a given variation in $\tau$ we will also introduce a
variation in the power spectrum amplitude 
that leaves the final $C_\ell$ unchanged. This is obtained by
multiplying the amplitude 
by a factor $\exp (a \delta \tau)$, where $a=2,3$ in the locl and
equilateral case respectively. A small shift $\delta \tau$ in the ionisation optical depth
then implies a shift in the amplitude equal to $\delta \A \simeq 2 \A \delta \tau$. 
The total bispectrum variation $\delta B$ is then given by (we
omit the subscript $\ell_1 \ell_2 \ell_3$ for 
simplicity of notation):
\beqarr
\frac{\delta B}{\delta \tau} \delta \tau & = & \frac{\partial
  B}{\partial \tau} \delta \tau 
+ \frac{\partial B}{\partial \A} \frac{\delta \A}{\delta \tau} \delta \tau \nonumber \\
         & = & -a B e^{-a \delta \tau} \delta \tau + 4 B e^{a \delta \tau} \delta \tau \nonumber \\
         & \simeq & (4-a) B \delta \tau \; ,
\eeqarr
where in the last line we neglected second order terms in $\delta \tau$. The total variation in $\fnlest$
for a given $\delta \tau$ is then:
\beq
\frac{\delta \fnlest}{\delta \tau} \delta \tau = (4-a) \fnlest \delta \tau
\eeq
The remaining parameter
to take into account is the scalar spectral index $n$. Our next step
is then the evaluation of ${\delta B / \delta n}$. First of all we
note that when changing 
$n$ we have to change the power
spectrum normalisation accordingly because the normalisation is
defined at a given pivot scale. To compute $\delta B$ arising from a small 
change in the spectral index we then have to evaluate again:
\beq
\frac{\delta B}{\delta n} \delta n = \frac{\partial B}{\partial n} \delta n + \frac{\partial
  B}{\partial A} \frac{\delta A}{\delta n} \delta n \; ,
\eeq
where the partial derivative with respect to $n$ is taken by assuming
$A$ fixed. 
The authors of \cite{Liddleetal} use
WMAP 3-years data to find that the normalisation is well fit by the following
expression:
\beq
\A^{WMAP} = \tilde{A}_S \frac{\exp(-1.24+1.04r)(1-n)}{\sqrt{1+0.53r}} \; ,
\eeq
where $r$ is the tensor-to-scalar ratio. We will use this ansatz,
with the additional assumption $r=0$. In this way we obtain, for
a given variation $\delta n$ of the scalar spectral index:
$\delta A = 1.24 A \delta n$ and, correspondingly, $\delta B \simeq
2.5 \delta n$. Finally, to approximately evaluate ${\partial B /
\partial n}$ we work in the pure SW regime. Estimates of the
signal-to-noise ratio have in this case been obtained by Komatsu and
Spergel \cite{KomatsuSpergel} for the equilateral configurations and
by Babich and Zaldarriaga \cite{BabichZaldarriaga} for the squeezed ones for $n=1$. Extending their results,  we obtain that in both cases:
\beq
\frac{\partial \fnlest}{\partial n} \simeq \frac{\fnl}{2}
\left[\log \left( \ell_{\rm max}
  \right) -\frac{1}{(1-n)} \right] \; ;
\eeq
this allows to write the variation in $\fnlest$ for a given $\delta n$
\beq
\frac{\delta \fnlest}{\fnlest} \simeq \left[2.5 -\frac{1}{2(1-n)} +\frac{1}{2} \log \left( \ell_{\rm max}
  \right) \right]  \delta n \; .
\eeq
Having an expression for the derivatives of the bispectrum 
with respect to each of the three parameters $\A$,$n$ and $\tau$ we
can now propagate the error
using Eq. (\ref{eqn:propagation}), that for this particular case
reads
\beq
\sigma_{\fnlest} = \sqrt{\left(\frac{\delta \fnlest}{\delta \tilde{A}_S}\right)^2 \sigma_{\tilde{A}_S}^2 +
\left(\frac{\delta \fnlest}{\delta \tau}\right)^2 \sigma_{\tau}^2+
\left(\frac{\delta \fnlest}{\delta n}\right)^2
\sigma_{n}^2} \; .
\eeq
Note that, as mentioned above, the correlation of $A_S$ with $\tau$
and $n$ is accounted for by the ansatz we have made for $A_S$:
\beq
A_S = \tilde{A}_S \exp\left[-a \tau -1.24(1-n)\right] \; ,
\eeq
whereas the correlation between $\tau$ and $n$ has been neglected in
the previous formula.
Using the expressions just derived above for ${\delta \fnlest / \delta
  \tilde{A}_S}$, ${\delta \fnlest / \delta \tau}$ and ${\delta \fnlest / \delta
  n}$  we finally get:
\beq 
\frac{\sigma_{\fnlest}}{\fnlest} = \sqrt{
  \left(2 \frac{\sigma_{\tilde{A}_S}}{A_S}\right)^2 
  + \left[ (4-a)\sigma_\tau \right]^2 
  + \left[2.5 -\frac{1}{2(1-n)} +\frac{1}{2} \log \left( \ell_{\rm max}
  \right) \right]^2
  \sigma_n^2} \; .
\eeq
The present WMAP 5-years analysis yields a fractional uncertainty on the
amplitude of the curvature power spectrum of order $3 \%$, while
$\sigma_\tau \simeq 0.016$ and $\sigma_{n} = 0.015$. Substituting
these numbers in the last formula yields a total fractional correction
of order $14 \%$ on $\fnlest$, in very good agreement with the
numerical results. If we now consider an experiment with the
characteristics of Planck, our Fisher matrix analysis give a
fractional uncertainty on $A_S$ of order $1.5 \%$, $\sigma_n 
= 0.004$, $\sigma_\tau = 0.005$. This produces a final fractional
correction of order $\simeq 5 \%$ on $\fnl$, again in very good
agreement with the numerical estimate.

To understand whether these corrections are negligible or not we have
now to
compare it with the 1$\sigma$ 
uncertainty $\Delta \fnl$ of the estimator, obtained with fixed
cosmological parameters. The WMAP analysis finds 
$\Delta \fnlloc \simeq 30$, $\Delta \fnleq \simeq 100$, and central
values $\fnlloc \simeq 60$ and $\fnleq \simeq 70$. Using the
fractional uncertainties above we get a contribution to the final
error bar from cosmological parameters uncertainties that amounts to
$\Delta \fnlloc \simeq 10$ and $\Delta \fnleq \simeq 9$.
The effect of propagating cosmological parameters uncertainties
can then be neglected for the equilateral shape, where the error bars
are larger, whereas it is
more important for the local shape 
(about $30 \%$ of the presently quoted error bar). 
If we now consider Planck, Fisher matrix based forecasts predict 
$\Delta\fnlloc \simeq 5$ and $\Delta \fnleq \simeq 60$. If we assume
the $\fnl$ central values found by the WMAP analysis we obtain a small effect
for the equilateral case, whereas $\delta \fnlloc \simeq 3$, i.e. of
the same order of magnitude as $\Delta \fnlloc$. This analysis then suggests that the
effect of propagating uncertainties in the cosmological parameters on the
final $\fnlloc$ error bar should be taken into account if large
central values of $\fnlloc$ are found with Planck. Note that a value
of $\fnlloc$ of order $60$ would mean a many $\sigma$ detection with
Planck. Correcting the error bar in order to account for error 
propagation effects would not change this result but it would on the
other hand modify the level of significance of such a detection.

Before concluding this section, we would like to stress again that the 
estimator of $\fnl$ currently employed in the analyses
fixes the cosmological parameters at their best-fit values. A way to
reduce the impact of the uncertainties on the parameters would be to 
perform a joint likelihood analysis in which the cosmological
parameters are allowed to vary and then marginalise over their
uncertainties. Obtaining a forecast of the final $\fnl$ error if this
approach is taken is the purpose of the next section.

\section{Fisher matrix}\label{sec:Fisher}
\begin{figure}[h]
\begin{center}
\includegraphics[height=0.6\textheight,width = 0.9\textwidth]{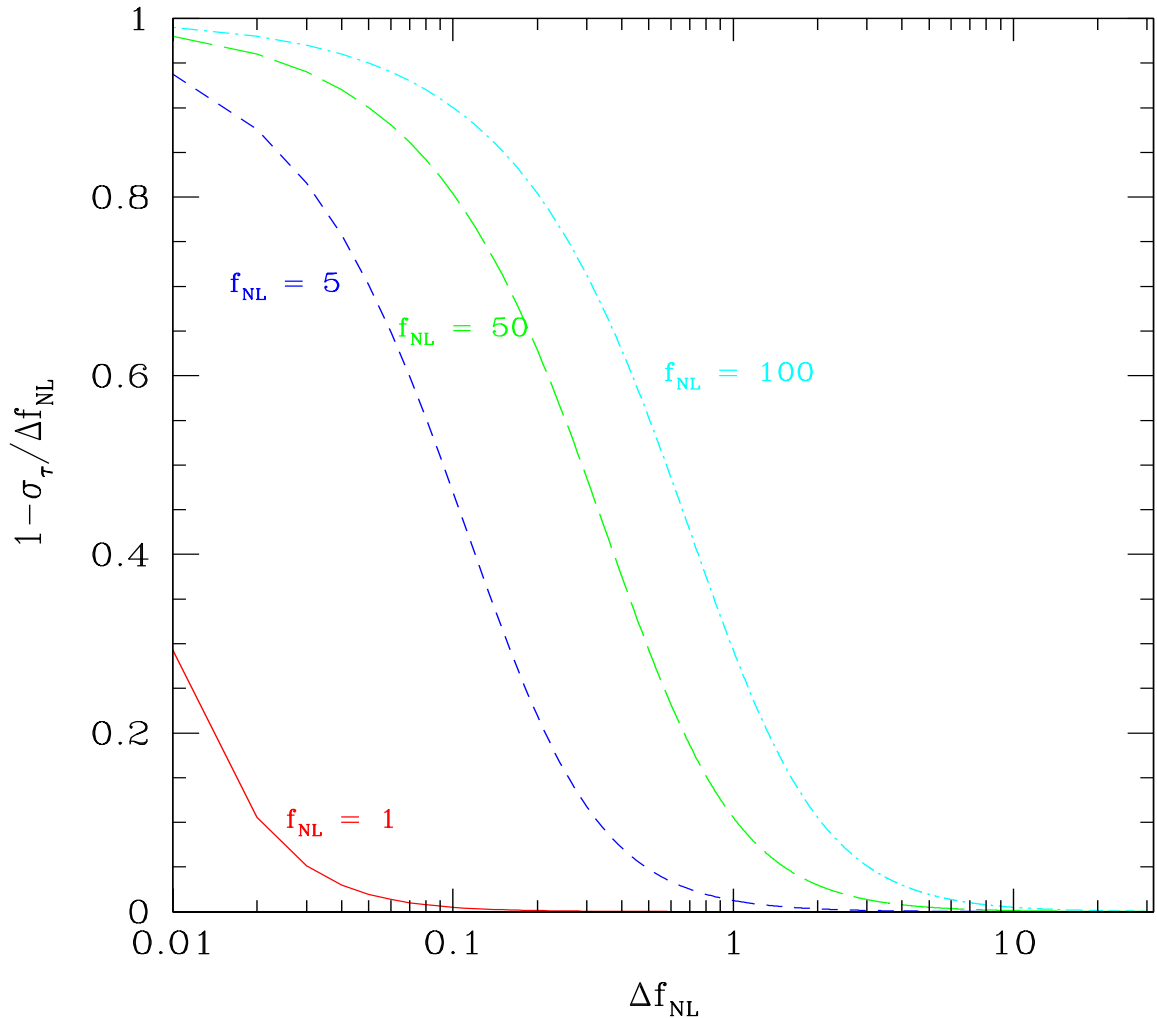}
\caption{Correction to $\Delta {\fnl}$ after marginalisation over
  $\tau$ is performed in a toy model where only uncertainties over $\tau$ are
  considered. We plot the correction as a function of $\Delta
  \fnl$. The correction becomes significant only when $\Delta \fnl$ is
  small enough to produce a many $\sigma$ detection for a given $\fnl$.  
}\label{fig:errtau}
\end{center}
\end{figure}

\noindent
As we were mentioning in the previous section, the optimal approach to
the $\fnl$
measurement would be to treat the cosmological parameters as nuisance
parameters and to marginalise over their distributions in order to get
the final $\fnl$ estimate. The
error on $\fnl$ can in this case be estimated by a Fisher matrix
analysis. If we consider a set
$\mathbf{p} = \{ p_i \}$ of cosmological parameters we can express the
Fisher matrix as \cite{KomatsuSpergel,SefusattiKomatsu}:
\beq\label{eqn:Fisher}
F_{ij} = \sum_{2 < \ell_1 < \ell_2  < \ell_3} \frac{\partial
  \Avbis}{\partial p_i} \frac{\partial \Avbis}{\partial p_j} \frac{1}{\sigma^2} \; ,
\eeq 
where $\sigma^2$ is the bispectrum variance. In the limit of small
non-Gaussianity we can take $\sigma^2 = C_{\ell_1} C_{\ell_2}
C_{\ell_3} \Delta_{\ell_1 \ell_2 \ell_3}$, where $\Delta$ takes the
values $1$,$2$,$6$ when two $\ell$'s are different, two of them are
the same and all are the same respectively. Following the results of
the previous
section, we know that the relevant set of parameters to consider is
$\mathbf{p} = \{\fnl, A, n, \tau\}$. Two account for cosmological
parameter uncertainties we  
add a Gaussian prior on the ith parameter with variance $\sigma_i^2$,
where $\sigma_i$ is the standard deviation obtained from the two-point 
function likelihood analysis. This approach is feasible as long as we
deal with weak non-Gaussianity and the two and three point function
can then be treated as uncorrelated.  
A Gaussian prior on the ith parameter with variance
$\sigma_i^2$ is imposed by simply adding a 
$\frac{1}{\sigma_i^2}$ term to the $ii$ entry of the Fisher matrix (see
e.g. \cite{TegmarkEfstathiou}). Once the Fisher matrix has been computed, 
the error on the $i$-th parameter after marginalising over
the others can be estimated in the standard way as:
\beq
\sigma_{p_i} = \sqrt{F_{ii}^{-1}} \; .
\eeq
Before moving to the numerical evaluation of formula
(\ref{eqn:Fisher}) for the full set of parameters 
let us start with a simplified case in which only $\fnl$ and $\tau$
are considered in the analysis and let us for simplicity restrict ourselves to the local
case. In this case, having made the approximation
(explained in the previous section) ${\partial B / \partial \tau} =
\exp(-2 \tau) B$, a simple analytical calculation gives
  the following Fisher matrix:
\begin{equation}
F = \left(\begin{array}{cc}
          \sum \frac{B_{\ell_1 \ell_2 \ell_3}^2}{\sigma_B^2}       
      &   -2f_{\rm NL} \sum \frac{B_{\ell_1 \ell_2 \ell_3}^2}{\sigma_B^2} \\
          -2f_{\rm NL}  \sum \frac{B_{\ell_1 \ell_2 \ell_3}^2}{\sigma_B^2}  
      &   4f_{\rm NL}^2 \sum \frac{B_{\ell_1 \ell_2 \ell_3}^2}{\sigma_B^2}
       \end{array}\right)
        \; ,
\end{equation}
where $\sigma_B^2$ is the bispectrum variance defined above.
This matrix is singular for $f_{\rm NL} \ne 0$, meaning that $f_{\rm NL}$ and $\tau$
are degenerate parameters. Adding a Gaussian prior on $\tau$
with variance $\sigma_\tau^2$ breaks the degeneracy.
\begin{equation}
F = \left(\begin{array}{cc}
          \sum \frac{B_{\ell_1 \ell_2 \ell_3}^2}{\sigma_B^2}       
      &   -2f_{\rm NL} \sum \frac{B_{\ell_1 \ell_2 \ell_3}^2}{\sigma_B^2} \\
          -2f_{\rm NL}  \sum \frac{B_{\ell_1 \ell_2 \ell_3}^2}{\sigma_B^2}  
      &   4f_{\rm NL}^2 \sum \frac{B_{\ell_1 \ell_2 \ell_3}^2}{\sigma_B^2}
          + \frac{1}{\sigma_\tau^2}
       \end{array}\right)
        \; .
\end{equation}
Inverting the Fisher matrix and taking the square roots yields the
final error on $\fnl$:
\begin{equation}
\sigma_{f_{\rm NL}} =
\sqrt{\frac{1}{\sum \frac{B^2}{\sigma_B^2}} 
                  \left( 1+4\sigma_\tau^2 f_{\rm NL}^2 
		  \sum
		  \frac{B^2}{\sigma_B^2} \right)} \; .	  
\end{equation}
If we call $\Delta f_{\rm NL}$ the estimated Fisher matrix error when we
{\em do not} marginalise over $\tau$ 
(i.e. the error usually quoted in the literature) then we see from the 
previous formula that:
\begin{equation}
\sigma_{\fnl} = \sqrt{(\Delta f_{\rm NL})^2 + 4 \sigma^2_{\tau} \fnl^2}
\; ;	  
\end{equation}
note the difference with respect to the previous approach in which
cosmological parameters were fixed. In that case the cosmological 
parameter errors biased the estimator and the uncertainties propagated
linearly (see also \cite{Creminellietal}):
\beq
\sigma_{\fnl} = \Delta {\fnl} + 2 \fnl \sigma_\tau \; .
\eeq
As we saw in the previous section, the error propagation scheme 
arising from the standard approach of fixing cosmological parameters
produces a relative correction of a few percent for WMAP and
Planck. We concluded that this correction is small but not always
negligible for Planck.
On the other hand the 
marginalisation approach used here makes the additional uncertainty 
much smaller than it was previously and always negligible, 
even for very large central value of $\fnl$, unless $\fnl$ is large enough
to produce a many $\sigma$ detection for a given experiment 
(see Fig. \ref{fig:errtau}).  

As long as only the parameter $\tau$ is considered we can then
conclude that both for WMAP ($\sigma_{\tau} = 0.016$, $\Delta \fnlloc
\simeq 30$) and Planck ($\sigma_{\tau} = 0.016$, $\Delta \fnlloc \simeq
5$) the correction to the $f_{\rm NL}$ error bars is totally negligible: 
$\delta \fnlloc < 0.2 \% \fnl$, assuming a central value $\fnl \simeq
60$. 
From the formula above we basically see that
the effect of marginalising over $\tau$ is to suppress the correction
mentioned in the previous section by a further factor
$\Delta {\fnl}$. Moreover we recover the correction mentioned in the
previous section in the limit $\Delta \fnl \rightarrow 0$. All this
makes sense: the correction from cosmological parameters uncertainties
is significant only if the error bar on $\fnl$ before
marginalisation is comparable to the error bars on the other
parameters; moreover a full likelihood estimation optimises 
the final error bar on $\fnl$ with respect to an analysis in which the
cosmological parameters are held fixed. The same results arise when we
account not only for $\tau$, but we consider the full set $\{A, n,
\tau, \fnl \}$. In this case we evaluated $\delta \fnl$ 
numerically from formula (\ref{eqn:Fisher}) and obtained that the
correction on the $\fnl$ error bar after marginalisation is always 
less than $0.5 \%$. 
The conclusion is that if  a full
likelihood analysis including the two and three point functions 
is applied in order to estimate $\fnl$, then the
impact of cosmological parameters uncertainties is totally
negligible. 

\section{Conclusions}\label{sec:Conclusions}
\noindent
In this paper we considered the effect of propagating cosmological 
parameters uncertainties on the estimate of the primordial
NG parameter $\fnl$. We firstly show that, accounting for
the large central value of $\fnl$ presently measured
\cite{YadavWandelt,WMAP5}, the final correction from parameters
uncertainties is of order $30 \%$ of the quoted $\fnlloc$ error bar for
WMAP and at about the same level of the predicted $\fnlloc$ error
bars for Planck. If a large $\fnl$ will be observed by Planck,
the effect of these uncertainties will then be not big enough to
change the conclusion that a large level of primordial non-Gaussianity
is present in the data. However the effect is important enough to change
the significance of the detection and should be taken into account
when quoting the error bars. We finally show that the effect of
cosmological parameters uncertainties becomes totally
negligible if we do not fix the cosmological parameters in the analysis,  but we 
treat them as nuisance parameters and marginalise over their
distribution in order to obtain the final $\fnl$ estimate. 
Even if optimal, this last approach is nevertheless
probably still inconvenient. A joint-likelihood 
evaluation would require a large amount of time and the final gain 
in the error bar would be significant only for large values of $\fnl$,
but those would produce a significant detection even in the 
sub-optimal approach. 

\acknowledgements
This work is partially supported by the European Community'sResearch 
Training Networks undr contracts MRTN-CT-2004-503369 and 
MRTN-CT-2006-035505.

{}

\end{document}